\documentclass[aps,twocolumn,superscriptaddress,prd,10pt]{revtex4-2}
\usepackage{graphicx}
\usepackage{amsmath,amssymb}
\allowdisplaybreaks[2]
\usepackage{hyperref}
\hypersetup{
  pdftitle=Tensor-polarized parton distribution functions of the deuteron by a convolution model,
  pdfauthor=Kenshi Kuroki,
  colorlinks=true,
  allcolors=blue,
}

\begin{document}
\title{Tensor-polarized parton distribution functions of the deuteron \texorpdfstring{\\}{}
by a convolution model}    
\author{S. Kumano}
\email{kumanos@impcas.ac.cn}
\affiliation
{Quark Matter Research Center,
    Institute of Modern Physics, Chinese Academy of Sciences,\texorpdfstring{\\}{}
    Lanzhou 730000, China}
\affiliation
{Southern Center for Nuclear Science Theory,
    Institute of Modern Physics, Chinese Academy of Sciences,\texorpdfstring{\\}{}
    Huizhou 516000, China}
\affiliation
{KEK Theory Center, Institute of Particle and Nuclear Studies, KEK,
Oho 1-1, Tsukuba 305-0801, Japan}
\author{Kenshi Kuroki}
\email{k-kuroki@impcas.ac.cn}
\affiliation
{Quark Matter Research Center,
    Institute of Modern Physics, Chinese Academy of Sciences,\texorpdfstring{\\}{}
    Lanzhou 730000, China}
\affiliation
{Southern Center for Nuclear Science Theory,
    Institute of Modern Physics, Chinese Academy of Sciences,\texorpdfstring{\\}{}
    Huizhou 516000, China}
\date{October 21, 2026}

\begin{abstract}
Tensor-polarized parton distribution functions (PDFs) 
are calculated for the deuteron by using a convolution formalism,
where the tensor-polarized PDFs are given by
the corresponding nucleon's unpolarized PDFs convoluted with 
the tensor-polarized nucleon momentum distribution in the deuteron. 
These distributions are obtained at $Q^2=2.5$ GeV$^2$ in order to compare 
with the tensor-polarized PDFs which were determined by HERMES $b_1$ data.
The obtained distributions are very different from the ones determined 
from the HERMES data, which indicates further studies are needed 
to clarify the difference, possibly by considering a new mechanism 
beyond the simple bound system of a proton and a neutron.
The obtained PDFs $\delta_T q $ and $\delta_T \bar q$ 
are converted to the PDFs of the Trento convention 
$f_{1LL}^{\, q}$ and $f_{1LL}^{\, \bar q}$, 
and they are used for estimating 
the twist-3 PDFs $f_{LT}^{\, q}$ and $f_{LT}^{\, \bar q}$
by using a Wandzura-Wilczek-like relation.
Because deep-inelastic-scattering experiments are under preparation 
for structure functions with a tensor-polarized deuteron target 
at the Thomas Jefferson National Accelerator Facility,
and a Drell-Yan experiment will be possible at hadron accelerator facilities,
such as the Fermi National Accelerator Laboratory,
the obtained tensor-polarized PDFs will be tested experimentally.
\end{abstract}
\maketitle

\section{Introduction}
\label{introduction}

High-energy spin physics has been investigated mainly for the spin-1/2 
nucleon to find the origin of the nucleon spin in terms of quarks 
and gluons by the fundamental theory of strong interactions,
quantum chromodynamics. 
By using spin-1 \cite{Kumano:2024fpr} and -3/2 \cite{Fu:2026mb}
hadrons and nuclei, it is possible to investigate different aspects 
of high-energy spin physics \cite{Frankfurt:1983qs,Hoodbhoy:1988am}. 
In this work, we focus on the spin-1 deuteron.
There was an experimental measurement by the HERMES Collaboration 
in 2005 for the tensor-polarized deuteron target 
\cite{Airapetian:2005cb}. 

There has been no such an experiment after 2005.
However, the situation is changing because 
the Thomas Jefferson National Accelerator Facility (JLab) is preparing
experiments with the tensor-polarized deuteron target 
to measure structure functions of the spin-1 deuteron
in the deep-inelastic-scattering (DIS) region \cite{Poudel:2025nof}. 
Furthermore, a Drell-Yan experiment is possible 
at hadron accelerator facilities, 
such as the Fermi National Accelerator Laboratory,
with the tensor-polarized deuteron target to investigate
tensor-polarized parton distribution functions (PDFs) \cite{Keller:2020wan}.
At the Nuclotron-based Ion Collider fAcility~\cite{Arbuzov:2020cqg}, the polarized deuteron beam is available 
for such studies. Similar experiments are also possible in future
by the Large Hadron Collider spin project \cite{Aidala:2019pit}
and at electron-ion colliders (EICs) 
\cite{AbdulKhalek:2021gbh,Anderle:2021wcy}.

It is especially encouraging that new data are expected from JLab, 
so that we are now stepping into a new field of 
high-energy spin physics, because the spin-1 deuteron contains 
structure functions which do not exist in the spin-1/2 nucleon.
Considering this situation of the experimental projects, 
we think that it is urgent to prepare theoretical formalism 
and predictions on the tensor-polarization observables.
In particular, the leading twist structure function $b_1$
will be measured first in the JLab experiment, and 
this function is expressed in terms of tensor-polarized
PDFs $\delta_T q$ and $\delta_T \bar q$.
In the Trento convention, these functions correspond to
$f_{1LL}^{\, q}$ and $f_{1LL}^{\, \bar q}$, respectively,
multiplied by a constant factor.
Standard theoretical predictions should be prepared for
$b_1$ and the tensor-polarized PDFs by using the current
knowledge of deuteron structure. 

Deuteron structure has been investigated mainly at low energies
for a long time. It is a bound system of a proton and a neutron
mainly in the S wave with the small admixture of the D wave.
This D-wave admixture gives rise to the tensor structure which
is observed as a finite electric quadrupole moment.
This is the standard model of the deuteron.
It is possible to calculate the structure function $b_1$
in this standard model by using a convolution-integral formalism,
which is generally used for calculating nuclear structure functions
\cite{Khan:1991qk,Hirai:2010xs}.
Its theoretical results were shown in Ref.\,\cite{Cosyn:2017fbo}
in comparison with the HERMES data. 
It is puzzling to find that the ``standard" deuteron estimates
are very different from the HERMES data, which may be 
interpreted by an exotic mechanism \cite{Miller:2013hla,Kaur:2025css},
for example, a hidden-color contribution.
Therefore, the tensor-polarized deuteron studies could lead to
a new field of hadron physics. 
There are related recent works in Ref.\,\,\cite{Tensor-SIDIS-2026}.

At this stage, the theoretical calculation is not shown
separately for the tensor-polarized valence-quark and 
antiquark distribution functions in the convolution model.
In future, all the tensor-polarized PDFs could be determined
from a global analysis of world data on tensor-polarized
measurements at high energies.
Namely, we expect that the tensor-polarized PDFs will be
shown for flavor-dependent valence-quark and antiquark distributions
and the gluon distribution.
In particular, it is the advantage of the Drell-Yan experiment
to find the antiquark distributions
\cite{Hino:1998ww,Hino:1999qi,Kumano:1999bt,Kumano:2016ude}.
For this purpose, it is good to show the flavor-dependent
valence-quark and antiquark distributions separately.
In addition, because the deviation from the $b_1$ sum rule \cite{Close:1990zw}
could be related to a finite tensor-polarized antiquark distribution,
such a separation should be valuable for discussing
possible tensor-polarized antiquark distributions in future.

In addition to the twist-2 tensor-polarized PDFs \cite{Bacchetta:2000jk},
it became possible to study tensor-polarized 
transverse-momentum-dependent parton
distribution functions (TMDs) and the PDFs up to twist 4
\cite{Kumano:2020ijt,Kumano:2021fem,Kumano:2021xau,Song:2023ooi}.
These TMDs will be investigated by semi-inclusive deep-inelastic-scattering (SIDIS) processes
\cite{Zhao:2025vol,Poudel:2025tac,Cosyn:2026vpc,Cosyn:2026cap}.
In the JLab experiments, higher-twist effects
could be sizable because $Q^2$ values are not very large
in comparison with a hadronic energy-scale squared. Here, 
$Q^2$ is given by the momentum transfer $q$ as $Q^2 = - q^2$.
Recently, twist-3 tensor-polarized distributions $f_{LT}$
were calculated \cite{Kumano:2025rai}
by using the twist-2 relation \cite{Kumano:2021fem,Kumano:2026xxv}
and the distributions $f_{1LL}$.
The twist-3 distributions $f_{LT}$ could be measured by the SIDIS, 
spectator-nucleon-tagging processes \cite{Cosyn:2020kwu}, 
and a Drell-Yan process \cite{Qiao:2024bgg}.

In this work, the convolution-model calculation is done
separately for the flavor-dependent valence-quark 
and antiquark distributions in the deuteron, 
and its numerical results are shown
for the tensor-polarized distributions 
$\delta_T q$ and $\delta_T \bar q$.
Then, they are compared with the tensor-polarized PDFs
\cite{Kumano:2010vz} determined from the HERMES data.
These distributions are converted into the Trento convention,
$f_{1LL}^{\, q}$ and $f_{1LL}^{\, \bar q}$,
which are used for calculating the twist-3 distributions 
$f_{LT}^{\, q}$ and $f_{LT}^{\, \bar q}$
by using the twist-2 Wandzura-Wilczek-like relation 
\cite{Kumano:2021fem,Kumano:2026xxv}.

This article consists of the following. 
First, the tensor-polarized structure functions and PDFs
are introduced in Sec.\,\ref{b1}, and 
the convolution formalism is explained 
in Sec.\,\ref{convolution}. Then, the numerical results
are shown in Sec.\,\ref{results}, and they are summarized
in Sec.\,\ref{summary}.

\section{Structure function \texorpdfstring{\boldmath$b_1$}{b1} and tensor-polarized PDFs}
\label{b1}

Because the tensor-polarized structure function $b_1$ and
the quark and antiquark distribution functions $\delta_T q$ and
$\delta_T \bar q$ are used in this work, they are introduced first
in the following. The charged-lepton deep inelastic cross section from
the spin-1 deuteron is given by the hadron tensor
\cite{Hoodbhoy:1988am,Kumano:2024fpr}
\begin{align}
W_{\mu \nu}^{\lambda_f \lambda_i}
 & =  -F_1 \hat{g}_{\mu \nu} 
     +\frac{F_2}{M \nu} \hat{P}_\mu \hat{P}_\nu 
     + \frac{ig_1}{\nu}\epsilon_{\mu \nu \alpha \beta} q^\alpha s^\beta 
\nonumber\\[-0.05cm]
& \hspace{0.37cm}
     +\frac{i g_2}{M \nu ^2}\epsilon_{\mu \nu \alpha \beta} 
       q^\alpha (P \cdot q \, s^\beta - s \cdot q \, P^\beta )
\nonumber\\[-0.05cm]
& \hspace{0.37cm}
  -b_1 r_{\mu \nu} + \frac{1}{6} b_2 (s_{\mu \nu} +t_{\mu \nu} +u_{\mu \nu}) 
\nonumber\\[-0.05cm]
& \hspace{0.37cm}
     + \frac{1}{2} b_3 (s_{\mu \nu} -u_{\mu \nu}) 
     + \frac{1}{2} b_4 (s_{\mu \nu} -t_{\mu \nu}) .
\label{eqn:Wmunu}
\end{align}
The tensors $r_{\mu \nu}$, $s_{\mu \nu}$, $t_{\mu \nu}$, 
and $u_{\mu \nu}$ are given by
\begin{align}
r_{\mu \nu} & = \frac{1}{\nu ^2}
   \bigg [ q \cdot E ^* (\lambda_f) q \cdot E (\lambda_i) 
           - \frac{1}{3} \nu ^2  \kappa \bigg ]
   \hat{g}_{\mu \nu}, \ \ \ 
\nonumber\\[-0.05cm]
s_{\mu \nu} & = \frac{2}{\nu ^2} 
   \bigg [ q \cdot E ^* (\lambda_f) q \cdot E (\lambda_i) 
           - \frac{1}{3} \nu ^2  \kappa \bigg ]
\frac{\hat{P}_\mu \hat{P}_\nu}{M \nu}, 
\nonumber\\[-0.05cm]
t_{\mu \nu} & = \frac{1}{2 \nu ^2}
   \bigg [ q \cdot E ^* (\lambda_f) 
           \left\{ \hat{P}_\mu \hat E_\nu (\lambda_i) 
                 + \hat{P} _\nu \hat E_\mu (\lambda_i) \right\}
\nonumber\\[-0.05cm]
& \hspace{0.30cm}
   + \left\{ \hat{P}_\mu \hat E_\nu^* (\lambda_f)  
           + \hat{P}_\nu \hat E_\mu^* (\lambda_f) \right\}  
     q \cdot E (\lambda_i) 
   - \frac{4 \nu}{3 M}  \hat{P}_\mu \hat{P}_\nu \bigg ] ,
\nonumber\\[-0.05cm]
u_{\mu \nu} & = \frac{M}{\nu} 
   \bigg [ \hat E_\mu^* (\lambda_f) \hat E_\nu (\lambda_i) 
          +\hat E_\nu^* (\lambda_f) \hat E_\mu (\lambda_i) 
\nonumber\\[-0.05cm]
& \hspace{0.30cm}
   +\frac{2}{3}  \hat{g}_{\mu \nu}
   -\frac{2}{3 M^2} \hat{P}_\mu \hat{P}_\nu \bigg ] ,
\label{eqn:rstu}
\end{align}
where $\hat{g}_{\mu \nu}$ and $\hat{a}_\mu$ are defined by
$\hat{g}_{\mu \nu} = g_{\mu \nu} - {q_\mu q_\nu}/{q^2}$ and
$\hat{a}_\mu = a_\mu - ({a \cdot q}/{q^2}) q_\mu $,
$\epsilon_{\mu \nu \alpha \beta}$ is 
the antisymmetric tensor with $\epsilon_{0123}=1$, 
$\nu$ is given by $\nu ={P \cdot q}/{M}$ 
with the spin-1 deuteron mass $M$ and its momentum $P^\mu$,
and $\kappa$ is defined by $\kappa= 1+{Q^2}/{\nu^2}$.
The index $\lambda_i$ ($\lambda_f$) is 
the initial (final) spin state of the deuteron.
The $E^\mu$ is the polarization vector to satisfy
the conditions $P \cdot E =0$, $E^2 = -1$, and 
\begin{align}
\sum_\lambda E^\mu (\lambda) E^{* \nu}  (\lambda)
    = - g^{\mu\nu} + \frac{P^\mu P^\nu}{M^2} .
\end{align}
Then, the spin vector of the deuteron is defined by
\begin{align}
(s_{\lambda_f \lambda_i})^{\mu}
    & = -\frac{i}{M} \epsilon ^{\mu \nu \alpha \beta} 
                E^*_\nu (\lambda_f) E_\alpha (\lambda_i) P_\beta .
\end{align}
In the rest frame of the deuteron, the polarization vector
is given by
\begin{align}
E^\mu (\lambda= \pm 1) & = \frac{1}{\sqrt{2}}(0,\mp 1, -i,0),
\nonumber \\
E^\mu (\lambda=0) & = (0,0,0,1) .
\end{align}

The $F_{1,2}$ are unpolarized structure functions 
and $g_{1,2}$ are polarized ones in Eq.\,(\ref{eqn:Wmunu}).
There are four structure functions $b_{1-4}$ which do not exist
in the spin-1/2 nucleon. These structure functions $b_{1-4}$
are associated with the tensor structure of the spin-1 deuteron.
Among them, twist-2 functions are $b_1$ and $b_2$,
and they are related with each other
by the Callen-Gross-like relation. The functions $b_3$ and $b_4$
are higher-twist ones. 
The tensors $r_{\mu \nu}$, $s_{\mu \nu}$, $t_{\mu \nu}$, 
and $u_{\mu \nu}$ in Eq.\,(\ref{eqn:rstu}) are symmetric
under the exchanges $\mu \leftrightarrow \nu$ and 
$E \leftrightarrow E^*$, so that $b_{1-4}$ can be measured
by an unpolarized charged-lepton beam.

In this work, the tensor-polarized PDFs are calculated
for the deuteron. The scaling variable is sometimes misleading
for nuclear structure functions and PDFs, so that
the variables $x$ and $Q^2$ are defined in the following.
The momentum transfer in the charged-lepton DIS is denoted as $q$.
For the deuteron, the Bjorken scaling variables are defined as
\begin{align}
x = \frac{Q^2}{2 M_N \nu}, \ \ \ 
x_{D} = \frac{Q^2}{2 P \cdot q} = \frac{Q^2}{2 M \nu} ,
\label{eqn:x-Bjorken}
\end{align}
where $M_N$ is the nucleon mass, 
and $\nu \, (=q^0$) is the energy transfer.
Their kinematical ranges are 
\begin{equation}
    \label{eqn:x-ranges}
    0 \le x \le 2, \ \ \ 
    0 \le x_{D} \le 1 .
\end{equation}
In showing the structure functions and PDFs of the deuteron,
the variable $x$ is usually used, so that this variable is used 
in the following.

The $b_1$ is expressed in terms of the tensor-polarized PDFs.
In the parton model or in the leading-order (LO) of the running coupling
constant $\alpha_s$, it is given by the tensor-polarized 
quark and antiquark distribution functions 
$\delta_{_T} q$ and $\delta_{_T} \bar q$ as
\begin{align}
b_1 (x,Q^2) & = \frac{1}{2} \sum_i e_i^2 
      \, \left [ \delta_{_T} q_i (x,Q^2) 
      + \delta_{_T} \bar q_i (x,Q^2)   \right ] ,
\nonumber \\[-0.05cm]  
       \delta_{_T} q_i 
        & = q_i^0  - \frac{q_i^{+1} + q_i^{-1} }{2} ,
\label{eqn:b1-parton}
\end{align}
where $i$ is the quark flavor, $e_i$ is the quark charge
in the unit of elementary charge, and
$q_i^\lambda$ is an unpolarized quark distribution
in the deuteron with the spin state $\lambda$.
The distribution $\delta_{_T} q$ is sometimes
denoted as $\delta q $ or as the Trento-convention PDF
$f_{1LL}^{\, q} \, [= - (2/3) \delta_T q ]$.

For estimating $b_1$, there is a useful sum rule
based on the parton model for the spin-1 deuteron
\cite{Close:1990zw},
\begin{align}
\int_0^2 \! dx \, b_1 (x,Q^2) 
    =  \sum_i e_i^2 \int_0^2 \! dx \, \delta_T \bar q_{i} (x,Q^2) .
\label{eqn:b1-sum}
\end{align}
If there is no tensor-polarized antiquark distribution,
the sum becomes $\int_0^2 dx \, b_1 (x)=0$.
If experimental data indicate a finite sum for $b_1$,
it could mean that a finite antiquark distribution exists.
The sum rule is similar to the Gottfried sum rule
\cite{Kumano:1997cy},
\begin{align}
\! \! \!
\int_0^1 \frac{dx}{x} & [F_2^p (x,Q^2)  - F_2^n (x,Q^2) ]
\nonumber \\
&    = \frac{1}{3} 
     +  \frac{2}{3} \int_0^1 \! dx \, [ \bar u(x,Q^2) - \bar d(x,Q^2) ] ,
\label{eqn:gottfried}
\end{align}
which is also based on the parton model. The deviation from 1/3 
indicates the finite asymmetric distributions $\bar u$ and $\bar d$.
The violation of the Gottfried sum initiated flourishing theoretical studies
on the physics origin of the asymmetric distribution $\bar u-\bar d$.
In the same way, the violation of the $b_1$ sum could indicate
an interesting hadronic mechanism.

According to the HERMES measurement in 2005, the sum is given as
$
\int_{0.02}^{0.85} dx \, b_1(x,Q^2)  =
   [0.35 $ $ \pm 0.10$ $ \text{ (stat)} \pm 0.18 \text{ (sys)}] \times 10^{-2} 
$
at $Q^2>1 \, \text{GeV}^2$ \cite{Airapetian:2005cb}.
At this stage, the experimental errors are large; however,
the finite sum could indicate a finite tensor-polarized antiquark distribution.
We expect to have JLab data in the near future, so that more accurate information
should be obtained for this sum. This situation motivated us to study
the tensor-polarized antiquark distributions based on a standard model
of the deuteron. 
In the next section, we explain how to calculate
the antiquark distributions in the deuteron as well as
the tensor-polarized quark distributions.

In addition to the twist-2 distribution functions, 
tensor-polarized higher-twist PDFs became possible 
to be investigated by recent theoretical progress
\cite{Kumano:2020ijt,Kumano:2021fem,Kumano:2021xau,Song:2023ooi,
Zhao:2025vol,Cosyn:2026vpc,Cosyn:2026cap,Kumano:2025rai,Kumano:2026xxv}.
The twist-3 distribution $f_{LT}$ can be written
in terms of the part expressed by the twist-2 contribution
$f_{LT}^{\,\text{twist-2}}$
and the dynamical (or genuine) twist-3 distribution $f_{LT}^{(HT)}$ as
\begin{align}
f_{LT} (x,Q^2) 
= f_{LT}^{\,\text{twist-2}} (x,Q^2)  + f_{LT}^{\,\text{(HT)}} (x,Q^2) .
\label{eqn:fltpm}
\end{align}
The twist-2 part can be calculated by
a useful Wandzura-Wilczek-like relation \cite{Kumano:2021fem,Kumano:2026xxv}
\begin{align}
f_{LT}^{\,\text{twist-2}} (x,Q^2) 
= \frac{3}{2} \int_x^2 \frac{dy}{y} f_{1LL} (y,Q^2) ,
\label{eqn:fltpm-twist2}
\end{align}
where the upper bound of the integral is 2 for the deuteron.
Therefore, as long as we neglect the dynamical twist-3 distribution
$f_{LT} ^{\,(HT)}$, we can estimate $f_{LT}$ 
by this twist-2 relation.
Another twist-3 PDF, the chiral-odd distribution $e_{LL}$, can also be decomposed into a quark-mass contribution and a genuine twist-3 contribution involving multiparton distribution functions~\cite{Kumano:2021xau}.
The equation-of-motion relation gives the quark-mass contribution in terms of the twist-2 distribution $f_{1LL}$ as $x e_{LL}^\text{mass}(x) = (m/M) f_{1LL}(x)$, where $m$ is the quark mass.
Since $m \ll M$, this contribution is expected to be small except possibly at very small $x$.
Therefore, a sizable $e_{LL}$ would be expected to arise predominantly from the genuine twist-3 contribution.

\section{Convolution description}
\label{convolution}

For calculating structure functions and PDFs of a nucleus,
a standard theoretical method is to use a convolution integral
of a structure function of the nucleon with a nucleon's 
momentum distribution in the nucleus.
For the details of the formalism, one may look at 
Refs.\,\cite{Hirai:2010xs,Cosyn:2017fbo}.
The structure function $b_1$ of the deuteron is given 
by the unpolarized structure function $F_1^N$ of the nucleon
and the tensor-polarized momentum distribution of a nucleon
in the deuteron $\delta_T f$ as
\begin{align}
b_1 (x,Q^2) & = \int^2_{x} \frac{dy}{y} 
\, \delta_T f(y) \, F_1^N (x/y,Q^2), 
\nonumber \\
\delta_T f(y) & = f^0 (y)  - \frac{f^{+1} (y) + f^{-1} (y)}{2} .
\label{eqn:b1-convolution}
\end{align}
Here, $f^\lambda (y)$ is the nucleon momentum distribution
in the deuteron spin state $\lambda$, and it is given by
the momentum-space wave function $\phi^\lambda (\vec p \,)$ as
\begin{align}
f^\lambda (y) = \int d^3 p \, y \, | \, \phi^\lambda (\vec p \,) \, |^2
          \, \delta \left ( y - \frac{E-p_z}{M_N}   \right ) .
\label{eqn:deuteron-momentum}
\end{align}
The variable $y$ indicates the light-cone momentum fraction
defined by
\vspace{-0.1cm}
\begin{equation}
y   =    \frac{M \, p \cdot q}{M_{N} \, P \cdot q} 
  \simeq \frac{2 \, p^-}{P^-} ,
\vspace{-0.1cm}
\end{equation}
where $p$ and $P$ are nucleon and deuteron momenta, 
respectively, and $p^-$ is the light-cone momentum defined by
$p^- = (p^0 -p^3)/\sqrt{2}$.
In the numerical analysis in Sec.\,\ref{results},
the nonrelativistic approximation is used for $p^0$,
and it is given by
$p^0 = M_N -\varepsilon - \vec p^{\; 2}/(2 M_N)$
with the separation energy $\varepsilon$ of the deuteron.

Expressing the wave function $\phi^\lambda (\vec p \,)$ 
in terms of S- and D-state wave functions
$\phi_0 (p)$ and $\phi_2 (p)$, we obtain the 
tensor-polarized nucleon momentum distribution as
\cite{Cosyn:2017fbo}
\begin{align}
\delta_T f(y)  = \int d^3 p \, y &
 \left [ - \frac{3}{4 \sqrt{2} \pi} \phi_0 (p) \phi_2 (p) 
  + \frac{3}{16\pi} |\phi_2 (p)|^2 \right ]
 \nonumber \\
 & \times
 (3 \cos^2 \theta -1) \, \delta 
 \left ( y - \frac{p\cdot q}{M_N \nu}  \right ) ,
\label{eqn:delta-t-f}
\end{align}
where $\theta$ is the polar angle of $\vec p$.
The normalization of the wave function is done 
by using the condition of the baryon-number conservation
$\int dy \, f^\lambda (y) 
= \int d^3 p \, y \, |\phi^\lambda (\vec p \,)|^2 =1$
\cite{Li:1988rj,Kumano:1989eh,Sargsian:2001gu,
CiofidegliAtti:2007ork,Hirai:2010xs,Geesaman:1995yd}.
This equation contains the extra factor $y$ in the integrand
and it is different from the nonrelativistic wave-function 
normalization. This normalization is used 
in the convolution formalism, and one may look at
original papers \cite{Li:1988rj,Kumano:1989eh,Sargsian:2001gu,
CiofidegliAtti:2007ork,Hirai:2010xs,Geesaman:1995yd}
on the details of this normalization.
The expression of Eq.\,(\ref{eqn:delta-t-f}) is slightly
different from the ones in Ref.\,\cite{Khan:1991qk}
as the details are explained in Ref.\,\cite{Cosyn:2017fbo}.

For calculating the convolution integral of Eq.\,(\ref{eqn:b1-convolution}), 
the nucleon's structure function $F_1^N$ is necessary.
It is calculated from $F_2^N$, which
is expressed by quark and antiquark distributions
in the LO and the longitudinal-transverse ratio 
$R=[ (1+Q^2/\nu^2) F_2^N -2x F_1^N]/(2xF_1^N)$ as
\begin{align}
& \! \! \! \! 
F_1^N (x,Q^2) = 
\frac{1+4 \, M_N^2 \, x^2/Q^2}{2 \, x \, [1+R(x,Q^2)]}
     \, F_2^N (x,Q^2) ,
\nonumber \\
& \! \! \! \! \! 
= \frac{1+4 \, M_N^2 \, x^2/Q^2}{2 \, [1+R(x,Q^2)]} 
\sum_i e_i^2 
     \left [ q_i^N (x,Q^2) + \bar q_i^N (x,Q^2) \right ] .
\label{eqn:f1-lo}
\end{align}
Here, the nucleon's distribution $q_i^N$ is defined 
by the distributions in the proton ($p$) and the neutron ($n$) as
$q_i^N = ( q_i^p + q_i^n ) /2$,
and the isospin symmetry is used as
$u^n = d^p \equiv d$, 
$d^n = u^p \equiv u$ and similar equations
for the antiquark distributions $\bar u$ and $\bar d$. 
Nuclear modifications exist for the longitudinal-transverse ratio $R$
for a nucleon in the deuteron as shown theoretically in 
Refs.\,\cite{Ericson:2002ep,Kumano:2025qzm}.
In addition, there are nuclear modifications in the PDFs themselves.
However, they are neglected in this work because
they are not large effects in the deuteron.
From Eqs.\,(\ref{eqn:b1-parton}), (\ref{eqn:b1-convolution}), 
and (\ref{eqn:f1-lo}), the convolution integral becomes
\begin{align}
& \sum_i e_i^2 \left[ \delta_T q_i^D (x,Q^2) 
                   + \delta_T \bar q_i^D (x,Q^2) \right ]
\nonumber \\
& \hspace{0.3cm}
= \sum_i e_i^2 \int^2_{x} \frac{dy}{y} \, \delta_T f(y)
\frac{1+4 \, M_N^2 \, (x/y)^2/Q^2}{1+R(x/y,Q^2)} 
\nonumber \\
& \hspace{2.3cm}
\times
\left[ q_i^N (x/y,Q^2) + \bar q_i^N (x/y,Q^2) \right ] .
\label{eqn:b1-convolution-2}
\end{align}

\begin{figure}[h!]
\centering
\includegraphics[width=4.0cm]{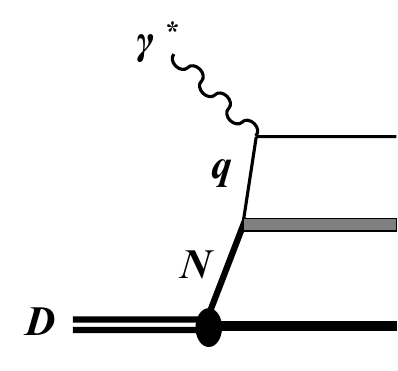}
\caption{Convolution model for the deuteron.
Quark and antiquark distribution functions of the deuteron are 
calculated by the corresponding functions of the nucleon convoluted 
with the nucleon's momentum distribution in the deuteron.}
\label{fig:convolution-1}
\end{figure}

Next, we assume that the convolution integral
is valid for each quark or antiquark
for calculating the tensor-polarized quark 
and antiquark distributions separately.
The tensor-polarized quark or antiquark distributions are
calculated by
\begin{align}
& 
 \left( 
    \begin{aligned}
      \, \delta_T q_i^D (x,Q^2) \, \\
      \, \delta_T \bar q_i^D (x,Q^2) \,
    \end{aligned}
 \right )
=  \int^2_{x} \frac{dy}{y} \, \delta_T f(y) 
\nonumber \\
& 
\hspace{1.2cm}
\times
\frac{1+4 \, M_N^2 \, (x/y)^2/Q^2}{1+R(x/y,Q^2)}
\left( 
    \begin{aligned}
      \, q_i^N (x/y,Q^2) \, \\
      \, \bar q_i^N (x/y,Q^2) \,
    \end{aligned}
 \right ) .
\label{eqn:b1-convolution-3}
\end{align}
Although Eq.\,(\ref{eqn:b1-convolution-2}) does not lead to
this relation uniquely, we consider that it is 
a reasonable first-step estimate on the tensor-polarized quark 
and antiquark distributions of the deuteron 
by the convolution model.
It means that each tensor-polarized quark or antiquark
distribution in the deuteron is simply given 
by the corresponding unpolarized distribution 
in the nucleon and the light-cone momentum distribution given by 
$\delta_T f(y)$ in the deuteron as illustrated
in Fig.\,\ref{fig:convolution-1}.

\section{Results}
\label{results}

\begin{figure*}[tbp]
    \centering
    \begin{minipage}[t]{0.49\hsize}
        \includegraphics[clip,width=0.95\linewidth]{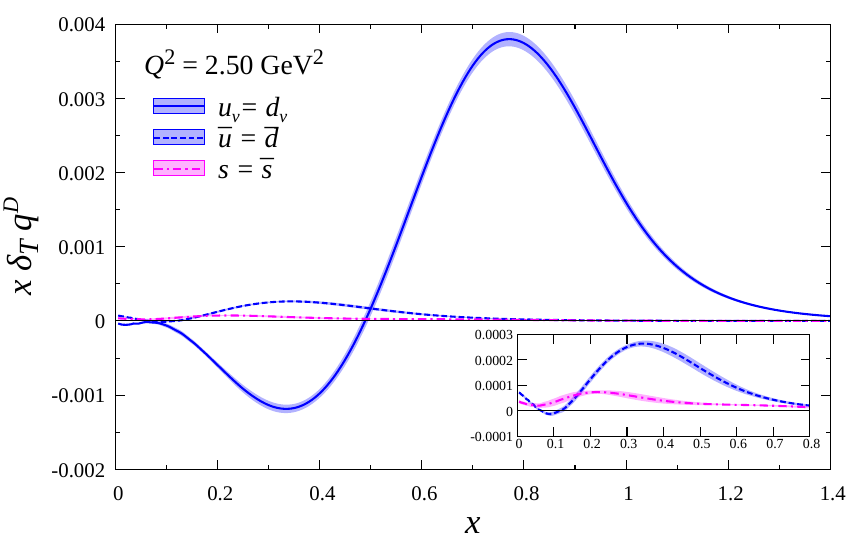}
    \end{minipage}
    \begin{minipage}[t]{0.49\hsize}
        \includegraphics[clip,width=0.95\linewidth]{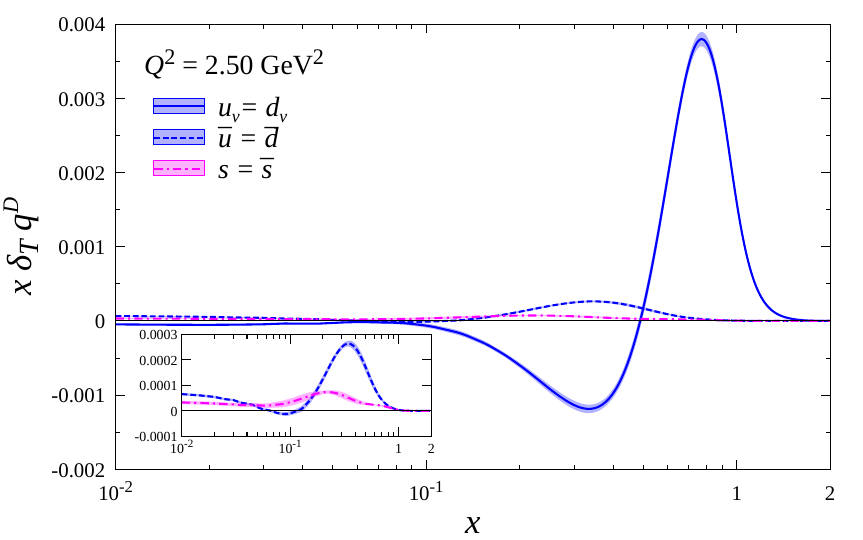}
    \end{minipage}
    \caption{Tensor-polarized quark and antiquark distributions by the convolution model at~$Q^2=2.5$~GeV$^2$. The distributions are shown with (left) a linear scale and (right) a logarithmic scale for $x$.
    The insets show enlarged views of the antiquark distributions.}
    \label{fig:deltaTq-conv_valence}
\end{figure*}

For the numerical analysis of Eq.\,(\ref{eqn:b1-convolution-3})
with the nucleon's momentum distribution in Eq.\,(\ref{eqn:delta-t-f}), 
the same functions are used as the ones in Ref.\,\cite{Cosyn:2017fbo}.
This work corresponds to the model 1 of Ref.\,\cite{Cosyn:2017fbo}.
For the deuteron wave function, the charge-dependent Bonn (CD-Bonn) function \cite{Machleidt:2000ge}
is used. As for the PDFs and the longitudinal-transverse ratio $R$ 
of the nucleon, the Martin-Stirling-Thorne-Watt, 2008 (MSTW2008) LO parametrization \cite{Martin:2009iq} 
and the SLAC-R1998 parametrization \cite{E143:1998nvx} are taken, respectively. 
For the strange- and antistrange-quark distributions, we use
$s^N = \bar s^N = (s +\bar s )_{\text{MSTW}}/2$.
Since the convolution formalism employed in this work is formulated at the leading order, we consistently use LO nucleon PDFs.
The separation energy of the deuteron is 2.22457 MeV \cite{Audi:2002rp}.
The dependencies of the calculated tensor-polarized PDFs on the nucleon PDFs and the deuteron wave functions are studied in Appendixes \ref{app:PDFs} and \ref{app:WFs}, respectively.

Calculated tensor-polarized valence-quark and antiquark distributions
are shown in Fig.\,\ref{fig:deltaTq-conv_valence} at $Q^2=2.5$~GeV$^2$.
Here, $D$ indicates the deuteron, and we denote it explicitly
in the following PDFs.
The error bands include the fit uncertainties of the nucleon PDFs at the 68\% confidence level~\cite{Martin:2009iq} and the fit uncertainty of the longitudinal-to-transverse ratio $R$~\cite{E143:1998nvx}.
The uncertainty of the deuteron wave function is not included, since no corresponding uncertainty estimate is available for the wave function used in this calculation.
The tensor-polarized valence-quark distributions $\delta_T u_v^D$ and
 $\delta_T d_v^D \, (=\delta_T u_v^D)$ are large at large $x$, 
and the antiquark distributions $\delta_T \bar u^D$,  
$\delta_T \bar d^D \, (=\delta_T \bar u^D$), and
$\delta_T \bar s^D \, (=\delta_T s^D)$ exist 
at relatively small $x$ and they are very small in comparison with 
the valence-quark distributions.
The different $x$ dependence of the valence- and sea-quark distributions originates from the convolution structure in Eq.~\eqref{eqn:b1-convolution-3}.
For a fixed deuteron momentum fraction $x$, the convolution samples the nucleon PDF at $x/y$ over the range $x \leq y \leq 2$.
Equivalently, by introducing $z=x/y$, the corresponding range of the nucleon momentum fraction is $x/2\leq z \leq 1$.
The valence distributions $q_i^N(z)$ are concentrated at moderate and large $z$, whereas the sea distributions $\bar{q}_i^N(z)$ become relatively more important toward smaller $z$.
Since the same tensor-polarized nucleon momentum distribution $\delta_T f(y)$ weights these different regions through the convolution, the resulting tensor-polarized valence- and sea-quark distributions acquire markedly different $x$ dependencies.
The convolution model is the standard way for calculating
nuclear structure function and PDFs, so that 
the calculated tensor-polarized PDFs are considered
as the ``standard" deuteron model predictions,
which should be tested by the PDFs determined
by a global analysis of world data in future.

\begin{figure}[b]
\centering
\includegraphics[width=0.95\linewidth]{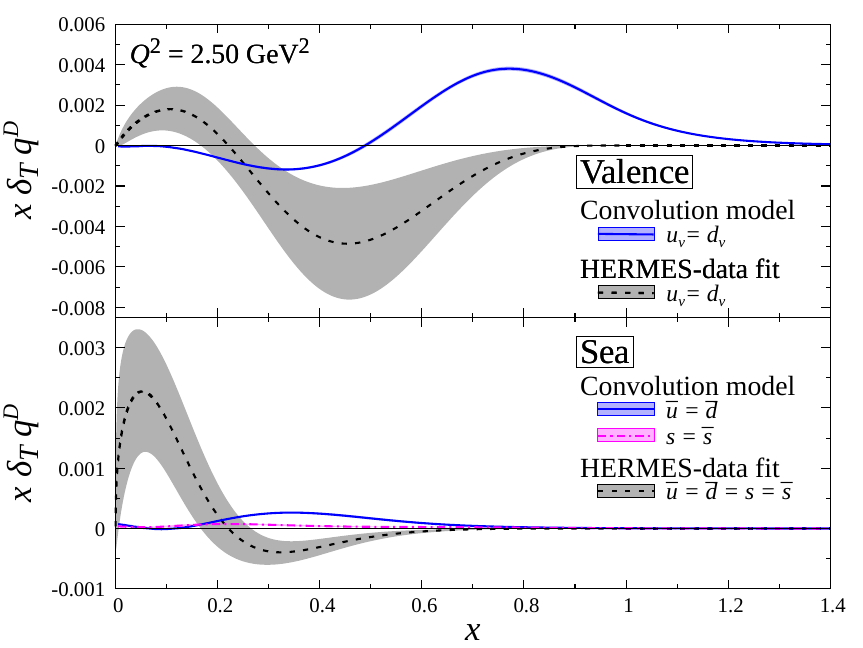}
\caption{Tensor-polarized quark and antiquark
distributions by the convolution model at~$Q^2=2.5$~GeV$^2$
in comparison with the PDFs to fit the HERMES data.}
\label{fig:tensor-polarized-PDFs-HERMES}
\end{figure}

At this stage, the HERMES data are the only existing ones, 
so that we may rely on the tensor-polarized PDFs determined 
by the fit to the HERMES data. 
In Fig.\,\ref{fig:tensor-polarized-PDFs-HERMES},
the calculated PDFs of this work are compared 
with the corresponding distributions obtained 
by a $\chi^2$ analysis of the HERMES $b_1$ data \cite{Kumano:2010vz}.
Our convolution-model distributions are very different
from the HERMES-data fit PDFs. 
The valence-quark distributions have different node locations
in $x$, and they have opposite oscillatory functional forms
in the sense that our distribution $\delta_T u_v^D \, (=\delta_T d_v^D)$
is negative at small $x$ ($<0.49$) and becomes positive 
at large $x$ ($>0.49$), where the HERMES-data fit 
$\delta_T u_v^D \, (=\delta_T d_v^D)$ is positive at small $x$ ($<0.22$) 
and it is negative at large $x$ ($>0.22$).
The antiquark distributions are also very different.
As noticed in Ref.\,\cite{Cosyn:2017fbo}, the convolution-model
results for $b_1$ are very different from the HERMES measurements, 
so that the differences in Fig.\,\ref{fig:tensor-polarized-PDFs-HERMES}
have the same issue.
At present, there is no established interpretation for the discrepancy. 
The errors of the HERMES data are relatively large, so that
we may wait for the JLab experiment for an independent 
experimental confirmation by accurate measurements. 
Then, it will become obvious whether the discrepancy
should be considered seriously,
possibly for finding a new hadronic physics.

\begin{figure}[b]
\centering
\includegraphics[width=0.95\linewidth]{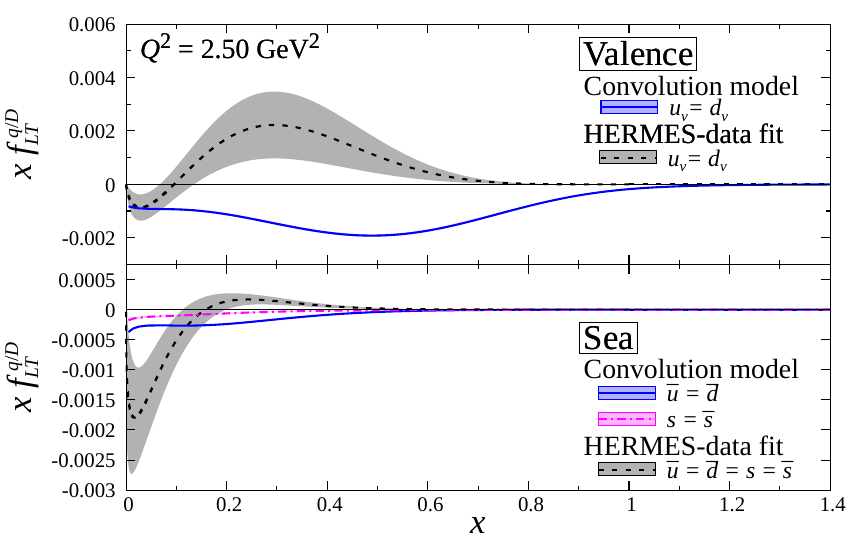}
\caption{Twist-3 tensor-polarized quark and antiquark
distributions $f_{LT}^{\, q/D}$ and $f_{LT}^{\, \bar q/D}$
calculated by using the Wandzura-Wilczek-like twist-2 relation
and the distributions $f_{1LL}^{\, q/D}$ and $f_{1LL}^{\, \bar q/D}$ 
obtained by the convolution model at~$Q^2=2.5$~GeV$^2$.
They are compared with $f_{LT}^{\, q/D}$ and $f_{LT}^{\, \bar q/D}$ 
calculated by using the twist-2 relation and $f_{1LL}^{\, q/D}$ 
and $f_{1LL}^{\, \bar q/D}$ obtained by the fit to 
the HERMES data \cite{Kumano:2025rai}.}
\label{fig:tensor-polarized-PDFs}
\end{figure}

Next, the tensor-polarized PDFs are converted to
the PDFs of the Trento convention by
\cite{Bacchetta:2000jk,Kumano:2024fpr} 
\begin{align}
f_{1LL}^{\, q/D} (x,Q^2) = - \frac{2}{3} \delta_T q^D(x,Q^2) .
\label{eqn:trento-pdfs}
\end{align}
The antiquark distributions are also calculated by this relation.
By the Wandzura-Wilczek-like twist-2 relation in 
Eq.\,(\ref{eqn:fltpm-twist2}) and the distributions
$f_{1LL}^{\, q/D}$,
the twist-3 distributions $f_{LT}^{\, q/D}$ are calculated 
by neglecting the dynamical twist-3 term as
\begin{align}
f_{LT}^{\, q/D} (x,Q^2) 
  = \frac{3}{2} \int_x^2 \frac{dy}{y} f_{1LL}^{\, q/D} (y,Q^2) ,
\label{eqn:fltpm-twist2-2}
\end{align}
and the same equation for the antiquark distributions.
The obtained distributions are shown 
in Fig.\,\ref{fig:tensor-polarized-PDFs}.
The twist-3 distributions themselves are of the order
of the twist-2 distributions, 
although their contributions to the cross section 
are suppressed by the $1/Q$ factor.
However, $Q^2$ values of the JLab measurements are
not large, so that the twist-3 effects could become 
sizable in the cross section and $f_{LT}^{\, q/D} (x)$
could be studied experimentally.

For example, 
the twist-2 part of $f_{LT}$ can be investigated indirectly in inclusive DIS through the leading-twist structure function $b_1$.
The $b_1$ data constrain the twist-2 distribution $f_{1LL}$, from which the twist-2 contribution to $f_{LT}$ can be estimated using the Wandzura-Wilczek-like relation~\cite{Kumano:2025rai}.
Beyond the leading twist, other tensor-polarized structure functions (e.g., $b_2$) exhibit direct sensitivity to twist-3 $f_{LT}$, and the tensor asymmetry $A_{zz}$, which incorporates these structure functions~\cite{Cosyn:2017fbo}, also directly includes the contribution of $f_{LT}$.
In addition to inclusive DIS, $f_{LT}$ could be measured in semi-inclusive 
deep inelastic scattering \cite{Zhao:2025vol}.
Its cross section is expressed by
the structure functions defined as
\vspace{-0.15cm}
\begin{equation}
F_\text{incoming lepton spin (target spin), photon polarization}
^\text{\ azimuthal angle} .
\nonumber
\vspace{-0.15cm}
\end{equation}
The twist-3 functions $f_{LT}^{\, q/D}$ and $f_{LT}^{\, \bar q/D}$ 
exist in $F_{U(LT)}^{\, \cos \phi_{LT}}$, which is obtained 
through $\phi_{LT}$ dependence of the cross section
for the unpolarized charged-lepton beam and the target with 
the tensor polarization (LT).
The $\phi_{LT}$ is the azimuthal angle 
of tensor polarization $S_{LT}^\mu$.
Another experimental possibility is to use Drell-Yan processes \cite{Qiao:2024bgg}.
The angular dependence of $\cos \hat\phi$ in the proton-deuteron Drell-Yan 
cross section can be used for finding $f_{LT}^{\, q/D}$ and 
$f_{LT}^{\, \bar q/D}$. Here, $\hat\phi$ is $\hat\phi = \phi-\phi_s$ with 
the azimuthal angle $\phi$ of $\mu^-$ and the transverse-vector angle $\phi_s$
of $S_{LT}^\mu$.

\section{Summary}
\label{summary}
The convolution model is the standard way for describing
nuclear structure functions. This model was used for calculating
the tensor-polarized twist-2 PDFs $\delta_T q^D$ 
and $\delta_T \bar q^D$, 
which were compared with the PDFs obtained by fitting the HERMES data
at $Q^2 =2.5$ GeV$^2$ for the deuteron.
Both PDFs are very different, which indicates that 
the tensor-polarized PDFs of the deuteron in this standard way 
are inconsistent with the HERMES data, although there may be
room due to higher-twist effects and experimental errors.
Then, the calculated twist-2 distributions and
Wandzura-Wilczek-like twist-2 relation were used for calculating
the twist-3 distributions $f_{LT}^{\, q/D}$ and $f_{LT}^{\, \bar q/D}$,
which could be measured experimentally.

\begin{acknowledgments}
The authors thank Qin-Tao Song for suggestions.
S.~K. and K.~K. thank the Chinese Academy of Sciences for its support.
K.~K. is also supported by the Gansu-province postdoctoral foundation.
\end{acknowledgments}

\appendix
\section{Dependence on the nucleon PDFs}
\label{app:PDFs}
The baseline calculation in this work uses the MSTW2008 LO PDF set~\cite{Martin:2009iq} because it was used in the previous convolution-model calculation~\cite{Cosyn:2017fbo}, allowing a direct comparison with the existing result. 
The convolution formalism employed here is formulated at the leading order, and therefore the LO PDF set is used consistently.
To investigate the dependence of the result on the choice of the nucleon PDFs, we additionally consider the MSHT20 LO~\cite{Bailey:2020ooq} and NNPDF4.0 LO~\cite{NNPDF:2021njg} sets.
MSHT stands for Mass Scheme Hessian Tolerance, and the MSHT20 is a subsequent PDF analysis in the MSTW2008 framework with updated data and parametrization, while NNPDF4.0 provides an independent PDF determination based on a neural-network analysis method.
For this comparison, the CD-Bonn deuteron wave function and the SLAC-R1998 longitudinal-transverse ratio are kept fixed, while only the nucleon PDF set is varied.
The different PDF sets are accessed using the LHAPDF6 library~\cite{Buckley:2014ana}.

\begin{figure}[tb]
\centering
\includegraphics[width=0.95\linewidth]{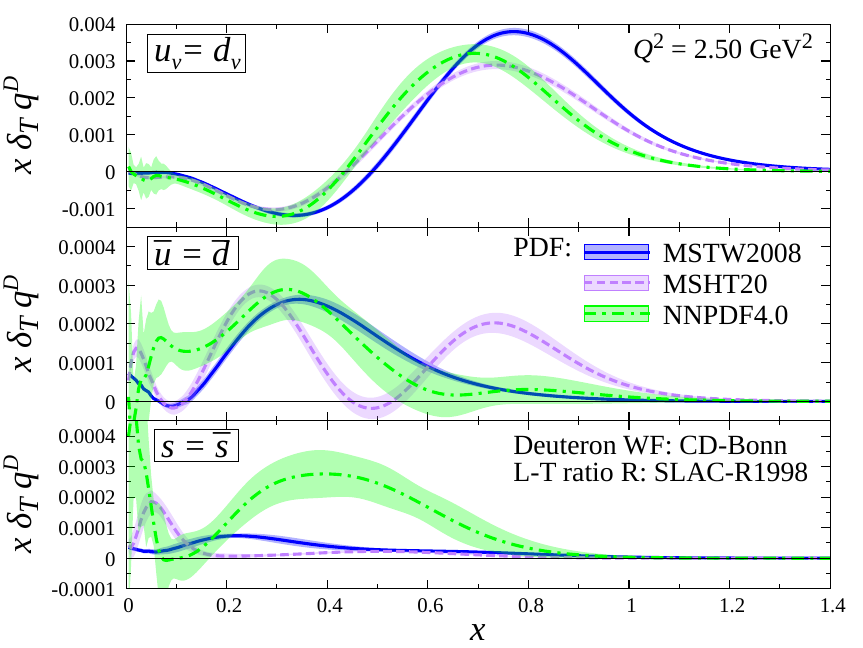}
\caption{Dependence of the tensor-polarized quark and antiquark distributions on the nucleon PDF set at~$Q^2=2.5$~GeV$^2$.
The results using the MSTW2008 LO, MSHT20 LO, and NNPDF4.0 LO PDF sets are shown.}
\label{fig:tensor-polarized-PDFs-PDFcompare}
\end{figure}

The resulting tensor-polarized deuteron PDFs are shown in Fig.~\ref{fig:tensor-polarized-PDFs-PDFcompare}.
The valence-quark distributions obtained with MSTW2008, MSHT20, and NNPDF4.0 are similar.
On the other hand, a larger dependence on the PDF set is observed for the sea-quark distributions, for which the PDF uncertainties are also substantially larger.
This reflects the weaker constraints on the sea-quark distributions in the nucleon compared with the valence distributions and is transmitted to the tensor-polarized deuteron PDFs through the convolution.
A particularly visible difference is found for the strange-quark distribution.
The NNPDF4.0 unpolarized strange-quark distribution is larger than those from MSTW2008 and MSHT20 in the relevant $x$ region.
Consequently, the tensor-polarized strange-quark distribution generated by the convolution is also larger when NNPDF4.0 is used.

\vspace*{0.2cm}
\section{Dependence on the deuteron wave functions}
\label{app:WFs}
The convolution-model prediction also depends on the deuteron wave function through the tensor-polarized nucleon momentum distribution $\delta_T f(y)$.
In the baseline calculation, the CD-Bonn wave function~\cite{Machleidt:2000ge} is used.
To examine the dependence on the nuclear input, we additionally consider the Nijm93~\cite{Stoks:1994wp} and Argonne v18~\cite{Wiringa:1994wb} wave functions.
The CD-Bonn potential is a relativistic nonlocal interaction, while Nijm93 and Argonne v18 are nonrelativistic local interactions.
For the numerical calculation, the momentum-space parametrizations of the Nijm93 and Argonne v18 deuteron wave functions given in Ref.~\cite{Zhaba:2016mtp} are used.
In this comparison, the MSTW2008 LO nucleon PDFs and the SLAC-R1998 longitudinal-transverse ratio are kept fixed, while only the deuteron wave function is varied.

\begin{figure}[b]
\centering
\includegraphics[width=0.95\linewidth]{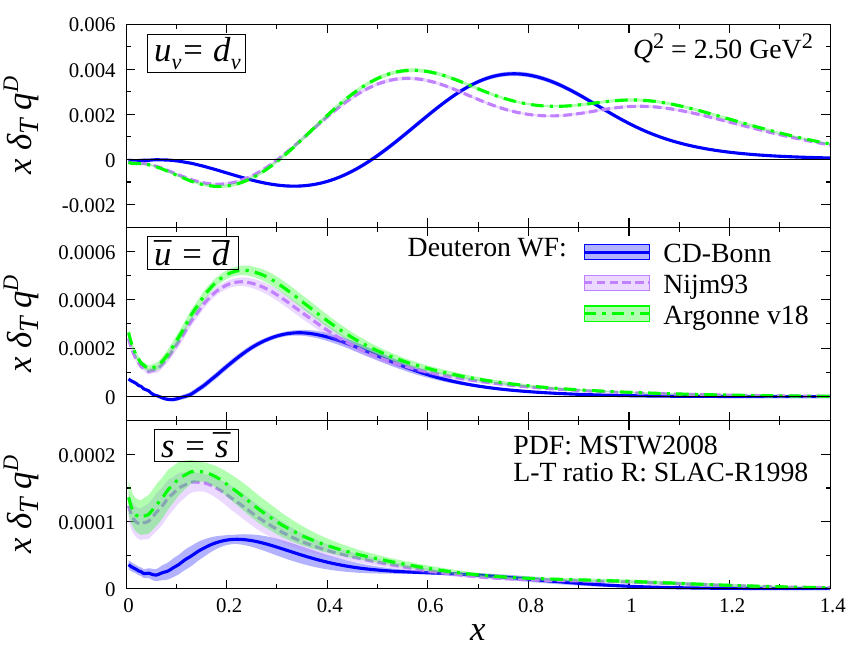}
\caption{Dependence of the tensor-polarized quark and antiquark distributions on the deuteron wave function at~$Q^2=2.5$~GeV$^2$.
The results using the CD-Bonn, Nijm93, and Argonne v18 wave functions are shown.}
\label{fig:tensor-polarized-PDFs-WFcompare}
\end{figure}

The results are shown in Fig.~\ref{fig:tensor-polarized-PDFs-WFcompare}.
The predictions obtained with Nijm93 and Argonne v18 are very close to each other.
In contrast, the CD-Bonn result is systematically smaller in magnitude.
This behavior is related to the different D-state components of the deuteron wave functions.
The D-state probabilities are $P_D=4.85\%$ for CD-Bonn and $P_D\sim5.76\%$ for Nijm93 and Argonne v18~\cite{Machleidt:2000ge,Stoks:1994wp,Wiringa:1994wb}.
Since the tensor-polarized nucleon momentum distribution in Eq.~\eqref{eqn:delta-t-f} is generated by the S-D interference and D-wave contributions, the smaller D-state component of CD-Bonn qualitatively leads to a smaller tensor-polarized momentum distribution.

\end{document}